\documentclass[runningheads]{llncs}
\usepackage{graphicx}
\usepackage[utf8]{inputenc}
\usepackage{balance} 
\usepackage{amsmath, nccmath}
\usepackage{array}
\newcolumntype{P}[1]{>{\centering\arraybackslash}p{#1}}
\usepackage{libertine}
\usepackage{amssymb}
\usepackage{dirtytalk}
\usepackage[autostyle]{csquotes}
\newcolumntype{L}{>{\centering\arraybackslash}m{10cm}}
\usepackage{comment}
\usepackage{subcaption}
\usepackage[export]{adjustbox}
\usepackage{wrapfig}
\usepackage{todonotes}
\usepackage{url}

\begin{document}
\title{A Bayesian model of information cascades}
%
%
\author{Sriashalya Srivathsan\inst{1} \and
Stephen Cranefield\inst{1} \and
Jeremy Pitt\inst{2}}
\authorrunning{S.\ Srivathsan et al.}
%
\institute{University of Otago, Dunedin, New Zealand
\email{ashal.srivathsan@postgrad.otago.ac.nz, stephen.cranefield@otago.ac.nz}\\
 \and
Imperial College, London, UK\\
\email{j.pitt@imperial.ac.uk}}
\maketitle              
\begin{abstract}
An information cascade is a circumstance where agents make decisions in a sequential fashion by following other agents. Bikhchandani et al.\ predict that once a cascade starts it continues, even if it is wrong, until agents receive an external input such as public information. In an information cascade, even if an agent has its own personal choice, it is always overridden by observation of previous agents' actions. This could mean agents end up in a situation where they may act without valuing their own information. As information cascades can have serious social consequences, it is important to have a good understanding of what causes them. We present a detailed Bayesian model of the information gained by agents when observing the choices of other agents and their own private information. Compared to prior work, we remove the high impact of the first observed agent's action by incorporating a prior probability distribution over the information of unobserved agents and investigate an alternative model of choice to that considered in prior work: weighted random choice. Our results show that, in contrast to Bikhchandani's results, cascades will not necessarily occur and adding prior agents' information will delay the effects of cascades.

\end{abstract}
\section{Introduction}

Propagation of opinions in society has a significant impact. In daily life it is clear that people are affected by others' views. For example, in electoral and financial campaigns, the spreading of news, opinion and rumours can have an enormous effect on the behaviour of the crowd. When people look at others' actions, or listen to others they update their assessment of the value of those actions and imitate accordingly.  

Information cascades are a social phenomenon in which all individuals from some point in a sequence onwards make the same decision. It occurs when other people's prior choices can strongly impact the choices of those who follow and it is the result of solely following others while discounting their own opinion. Bikhchandani et al.~\cite{bikhchandani_theory_1992} say that, \say{An informational cascade occurs when it is optimal for an individual, having observed the actions of those ahead of him, to follow the behavior of the preceding individual without regard to his own information}. This phenomenon can be observed, for example, when people choose a restaurant or school \cite{banerjee_simple_1992}.

People are more attracted when they see that someone already adopted a trend or chosen a specific fashion. New communication technology such as social media has become a habit, helping to share new trends, fashion and ideas \cite{cheng_can_2014}. It passes very quickly across web technologies when someone shares a particular fashion he or she likes, and this is subsequently shared by friends~\cite{lu_exploring_2020,liu_cascade_2020}. Resharing posts develops an information cascade in social media and leads others to choose a product, movie, or a specific fashion. Furthermore, recent studies~\cite{duan_informational_2009,liu_interaction_2019} have shown that  e-marketing is driven by information cascades. Customer ratings and  reviews make a huge impact in online shopping.  Thus, often people become more aware of and start to follow other's opinions when they need to shop online, rather than following their own preferences. It is evident that when shopping in Amazon or Ebay, people  rely heavily on others' choices since the quality of the goods is still an issue \cite{wang_why_2014}. 

In addition, an information cascade can aid society to unite its members for environmental collective action such as minimising emission of carbon and global warming as well as social and political collective action. Lohmann \cite{lohmann_collective_2000} notes that cascades are observed in society when individuals join and protest against a regime to address their political concerns. 

It is obvious that people often avoid their own preference or personal signals and choose to pursue others' choices as soon as a cascade begins. However, even though they have strong qualities, people's own preferences are concealed once the cascade starts. For instance, the algorithm for discovering the real rating of an Amazon product was proposed by Wang et al.~\cite{wang_why_2014} as people demand plausible and accurate online reviews. Wang et al.\ attempted as much as possible to eliminate the herding effect occurring through online purchasing using their algorithm. But the actual rating was very difficult to obtain.

Bikhchandani et al.~\cite{bikhchandani_theory_1992} have presented a probabilistic model to explain how an information cascade comes about, in the context of uptake of a fashion or fad. Their model predicts that cascades are inevitable, once they start. It is assumed that there is a true value of the fashion (0 or 1), which may be perceived correctly (with some probability $p>=0.5$) or incorrectly. A cascade can be either a \emph{correct} cascade, where eventually all individuals make the choice that aligns with the true value, or an \emph{incorrect} one, where eventually all choices go against the fashion's true value. The model uses a deterministic model of choice, which means that the agents choose the action (to adopt or reject the fashion) with the greatest evidence for its correctness, based on observations of other agents and their own private information, or make a 50/50 random choice if the evidence for \say{adopt} and \say{reject} is equal.

Bikhchandani et al.\ presented a high level analysis of their model, which lacks mathematical details. Our aim is to provide a detailed Bayesian model that accounts for the uncertainty about the private information of other agents. We also wish to investigate the impact of agents making choices non-deterministically, via random weighted choice. Indeed, this probabilistic choice model is claimed to be psychologically more realistic~\cite{luce_preference_1965,pleskac_decision_2015}.

In the model of Bikhchandani et al., the first agent in the chain of observed agents is assumed to be the first to make a decision to adopt or reject the fashion. This makes its choice highly influential, as its action is known to correspond directly to its private signal. Moreover, this is unrealistic in many settings. An agent may know that the fashion has been around for a while, so it could estimate a number of prior agents whose actions were unobserved. Our model therefore introduces uses a prior probability distribution over the count of positive perceptions of the fashion by these unobserved agents. Even if this is a completely uninformative prior distribution, it weakens the dominance of the first agent.

\section{Prior Work}
Besides Bikhchandani et al., many other researchers have  highlighted the importance of information cascades.
Benerji et al.~\cite{banerjee_simple_1992} introduced a model to investigate herd behaviour to understand how people adopt others' actions while ignoring their own information. They showed that people observe others' actions and tend to act in the same way because they believe the previous people have better information than them. Easley et al.~\cite{easley_networks_nodate} presented a theoretical and experimental study of herding behaviour. They presented a Bayesian model for sequential decision-making where people consider the counts of the previous actions and choose the highest one.

Vany et al.~\cite{vany_information_nodate} developed an agent-based model to compare the theoretical aspects of Bikichandani's concept. Specially, they analysed how people are attracted to watch a specific movie as they have multiple choices. The key aspect of this model is how an agent will be highly influenced by the nearest neighbour's action as well as the popularity of different movies. In the same fashion, Lee et al.~\cite{lee_i_2015} examined how online  reviews of a movie are socially influenced \cite{peng_social_2017} by two different groups such as a general crowd and friends. They concluded that the reviews of the popular movies show a cascading behaviour. However, people tend to follow friend ratings regardless of popularity, since these have more impact than the general crowd rating. Similarly Liu et al.~\cite{liu_influence_2020,liu_information_2014} analysed how information cascades occur in e-book marketing. Particularly, for both paid and free e-books they experimented to find the effect of information cascades, and found that an information cascade has a strong impact when selecting paid e-books in comparison with free books.

Anderson et al. \cite{anderson_information_2020}  designed a laboratory experiment to show how an information cascade occurs. It became apparent that people at some stage discard their own knowledge and continue to follow others. Huber et al.~\cite{huber_neural_2015} used functional magnetic resonance imaging of experimental participants' brains to research how a cascade can be stimulated by individual preference. They concluded from their findings that overweighting personal information can trigger and stop the information cascade.

 Watts \cite{watts_influentials_2007,watts_simple_nodate} developed a model to show how agents' actions  interact with neighbours' actions by setting a simple threshold rule. Cheng et al.~\cite{cheng_can_2014} addressed the prediction of a cascade when sharing a picture post in Facebook. They found that the size of an information cascade can be predicted by its temporal (observed time) and structural (caption, language and content) features. Lu et al.~\cite{lu_exploring_2020} proposed a system to analyse the collective behaviour of information cascades by analysing a huge social media data set.

\section{The information cascade model of Bikhchandani et al.}

Bikhchandani et al.~\cite{bikhchandani_theory_1992} considered a sequence of individuals who will choose between accepting ($A$) or rejecting ($R$) a fashion or fad. The model assumes that there is some true value $V \in \{0,1\}$ representing the benefit of following the fashion/fad. Every individual perceives a private signal $X_i \in \{H, L \}$, representing a \say{high} or \say{low} perception of $V$. As shown in Table \ref{tab:table1} a correct perception of $V$ has probability $p$ for $V = 1$ and $1\!-\! p$ for $V = 0$, and vice versa for incorrect perceptions. 

In the \say{basic model} considered by Bikhchandani et al., each agent considers the actions of earlier agents in the sequence, and takes those actions of adopt or reject as a proxy for what each of those agents perceived as a true value, which is unknown by observing agents. The agent can count all of the previous accepts or rejects, adds a count 0 or 1 (respectively) given his own information ($L$ or $H$), and then chooses his action based on the action with the greatest count. Here we notice that the actions of agents earlier in the chain are repeatedly used as evidence by later agents, and their impact may therefore be exaggerated. 

The \say{general model} of Bikhchandani et al.\ states that an agent will make its decision based on the expected value of $V$ given the agent’s private signal and observations of other agents’ actions. Bikhchandani et al.\ gave the following definition: $V_{n+1} (x; A_n)  \equiv E[V|X_{n+1} = x, X_i \in J_i(A_{i-1}, a_i) \text{, for all } i \leq n]$. Here $x$ is the signal perceived by agent $n+1$, $A_{i-1}$ is the sequence of prior actions that agent $i$ has observed and $J_i(A_{i-1}, a_i)$ is a set of signals that led individual $i$ to choose action $a_i$. Individual $n+1$ adopts if $V_{n+1}(x;A_n) \geq C$, where $C$ is the cost of adoption\footnote{This cost isn't used in our model.}. However, no computational account is given of how this expected value is determined by the agent.

Bikhchandani et al.\ found that \say{An informational cascade occurs if an individual's action does not depend on his private signal}. Hence, he ignores his private signal and will adopt based on the prior agents' actions alone.
This is true for all subsequent agents too, so they follow their predecessors and create a cascade. If $V=1$, a cascade can be either a correct cascade, where all adopt, or an incorrect cascade, where all reject, and vice versa for $V = 0$. 

Bikhchandani et al.\ showed that once the cascade starts it will last forever even if it is incorrect.  They also discussed the fragility of cascades. For instance, cascades can be broken if public information is revealed. 

\begin{figure}[t]
\centering
\includegraphics[scale= 0.5]{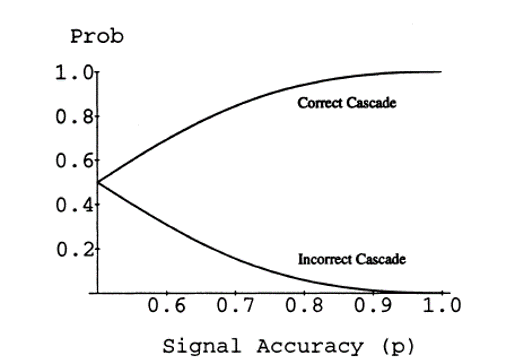}
\caption{Graph of probability of correct and incorrect cascade as a function of \textit{p} in the simple model of Bikhchandani et al.~\cite{bikhchandani_theory_1992}}
\label{fig:cascade}
\end{figure}
Our aim is to create a Bayesian model of an information cascade that captures the uncertainty about prior agents' private information. We avoid the strong influence of the first agent's action on later agent's action  through the ability to model a probability distribution over the information of unobserved prior agents. The model of Bikhchandani et al.\ uses a deterministic model of choice in which an agent ranks options and chooses the highest ranked one. This deterministic technique\footnote{Luce refers to this as \emph{algebraic choice} due to the common use of algebraic rather than probabilistic models in economics~\cite{luce_preference_1965}.} is commonly used in economic models \cite{blavatskyy_dual_2020}. Inspired by the work of Luce, \cite{pleskac_decision_2015,luce_preference_1965} we consider the effects on cascades if agents use weighted non-deterministic choice when choosing to adopt or reject. This means that choice is probabilistic and an agent may choose randomly from a set of weighted choices, in proportion to their weights.
This has been claimed to be more psychologically plausible than deterministic choice~\cite{pleskac_decision_2015}.

\begin{table}[t]
 \centering
    \begin{center}
\caption{Signal probabilities \cite{bikhchandani_theory_1992}}\label{tab:table1}
\begin{tabular}{|c|c|c|}
\hline
Gain of adopting &  $P(X_i = H \mid V)$ & $P(X_i = L \mid V)$\\
\hline
$V = 1$ &  $p$ & $1 - p$\\
$V = 0$ &  $1 - p$ & $p$\\
\hline
\end{tabular}
\end{center}
\end{table}

\section{Our Model and Approach}
\label{sec:model}
\begin{figure}[t]
  \centering
  \includegraphics[width=0.6\linewidth]{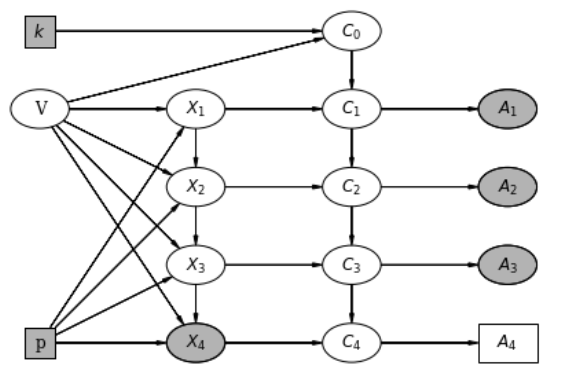}
  \caption{Probabilistic graphical model of information cascade from agent 4's viewpoint}
  \label{fig:Centipede}
\end{figure}

In this section we present our Bayesian model of information cascades. We define the following variables:
\begin{itemize}
    \item  $V \in \{ 0, 1 \} $, is the true value of the fad/fashion.
    \item  $X_i \in \{ H, L \} $, is the private signal of the agent: high or low. This represents agent $i$'s  possibly incorrect perception of $V$.
    \item  $A_i \in \{ A, R \} $, is the action of the agent: adopt or reject. 
    $C_i$ is a count of  how many times $H$ appears in \{$ X_1, \cdots, X_n$\}. This will be probabilistically inferred by each agent since it cannot  be directly perceived.
    \item $k$ is a number of prior agents who made choices that were not observed by any of the agents. $C_0$ is the count of $H$ signals observed by these prior agents. As this is unknown, our model uses an estimated probability distribution for $C_0$. 
    \item $p$ is the signal accuracy, which is assumed to be the same for all agents.
\end{itemize}
The dependencies between these variables are shown using a probabilistic graphical model \cite{pattern_recognition} in Figure \ref{fig:Centipede}, which is expressed from the viewpoint of agent 4. $k$ and $p$  are constants, but the other nodes are random variables. Shaded nodes represent observed variables, and $X_i$ is known by agent $i$. $A_4$ is shown as a rectangle, as this is a decision node. Agent 4 will choose $A$ if $P(V = 1 \mid X_4, A_1, A_2, A_3) > 0.5$, $R$ if this probability is less than 0.5, and otherwise will make a 50/50 choice between $A$ and $R$. An equivalent choice procedure involving $C_i$ is given at the end of this section.

The information cascade model is used as follows.
V and $C_i$ are conditionally independent given $C_{i-1}$ and $X_i$, therefore:
\begin{equation}
P(V, C_i \mid C_{i-1}, X_i) = p(V \mid C_{i-1}, X_i) \, p(C_i \mid C_{i-1}, X_i)
\end{equation}
Each agent $i$ maintains a joint probability distribution over $V$ and $C$ that is conditional on the actions observed so far. This is $p(V,C_0) = p(V)\, p(C_0 \mid V)$ for the first agent. We use a uniform prior over $V$ and a binomial distribution for $C_0$ given $V$: \begin{equation}
P(C_0=c \mid V=0) = (1-p)^c p^{k-c} \text{\ \ and \ \ } P(C_0=c \mid V=1) = p^c (1-p)^{k-c}
\end{equation}

For $i>1$, agent $i$ will observe the actions of all prior agents $j<i$ and compute the joint conditional distribution $p(V,C_j \mid C_0, A_1, \,\cdots\!, A_j)$. Agents act and are observed sequentially, so it will already have computed $p(V,C_{j-1} \mid C_0,A_1, \cdots\!, A_{j-1})$.\footnote{$C_0$ appears as a condition whenever a sequence of action variables does. Henceforth, we omit it for brevity.} Once it observes $A_j$, it first uses Bayes' Theorem to compute $P(X_j \mid A_1,\cdots,A_j):$ 

\begin{equation}
P(X_j \mid A_1,\cdots,A_j) \propto P(X_j \mid A_1,\,\cdots,A_{j-1}) \, P(A_j \mid X_j, A_1,\cdots,A_{j-1}) 
\end{equation}
\noindent where:
\begin{equation}
\begin{aligned}
 P(X_j \mid &A_1,\cdots,A_{j-1})\\
            &= \sum_{v\in\{0,1\}} P(X_j \mid V=v, A_1,\cdots,A_{j-1})\, P(V=v \mid A_1,\cdots,A_{j-1})\\
     &= \sum_{v\in\{0,1\}} P(X_j \mid V=v)\, P(V=v \mid A_1,\cdots,A_{j-1}) \\
\end{aligned}
\end{equation}
\begin{fleqn}
\begin{equation}
     P(A_j \mid X_j, A_1,\cdots,A_{j-1}) = P(A_j \mid X_j, C_{j-1})
\end{equation}
\end{fleqn}

\noindent $P(A_{j} \mid X_{j}, C_{j-1})$ is calculated as shown in Tables \ref{tab:table2} and \ref{tab:table3} for the deterministic and non-deterministic models of choice, respectively.

\begin{table}[t]
    \centering
    \begin{center}
    \caption{$P(A_{j} \mid X_{j}, C_{j-1})$ for deterministic choice}\label{tab:table2}
 \begin{tabular}{|c|c|c|c|} 
 \hline
 $X_{j}$ & Condition on $C_{j-1}$ & $P(A_j\!=\!A)$ & $P(A_j\!=\!R)$ \\ 
 \hline
      &  $P(C_{j-1} + 1 > \tfrac{k+j}{2}) > P(C_{j-1} + 1 < \tfrac{k+j}{2})$ & 1 & 0 \\[3pt] 
  $H$ &  $P(C_{j-1} + 1 > \tfrac{k+j}{2}) < P(C_{j-1} + 1 < \tfrac{k+j}{2})$ & 0 & 1 \\[3pt]
      & $P(C_{j-1} + 1 > \tfrac{k+j}{2}) = P(C_{j-1} + 1 < \tfrac{k+j}{2})$ & 0.5 & 0.5\\[3pt]
\hline
      &  $P(C_{j-1} > \tfrac{k+j}{2}) > P(C_{j-1} < \tfrac{k+j}{2})$ & 1 & 0 \\[3pt] 
  $L$ &  $P(C_{j-1} > \tfrac{k+j}{2}) < P(C_{j-1}  < \tfrac{k+j}{2})$ & 0 & 1 \\[3pt]
      & $P(C_{j-1} > \tfrac{k+j}{2}) = P(C_{j-1} < \tfrac{k+j}{2})$ & 0.5 & 0.5\\[3pt]
 \hline
\end{tabular}
\end{center}
\end{table}

\begin{table}[t]
    \centering
    \begin{center}
    \caption{$P(A_{j} \mid X_{j}, C_{j-1})$ for non-deterministic choice}\label{tab:table3}
 \begin{tabular}{|c|c|c|} 
 \hline
 $X_{j}$ & $P(A_j\!=\!A)$ & $P(A_j\!=\!R) $  \\ 
 \hline
  $H$ & $P(C_{j-1} + 1 > \tfrac{k+j}{2}) + \tfrac{1}{2} P(C_{j-1} + 1 = \tfrac{k+j}{2})$
      & $P(C_{j-1} + 1 < \tfrac{k+j}{2}) + \tfrac{1}{2} P(C_{j-1} + 1 = \tfrac{k+j}{2})$  \\[3pt]
 \hline
  $L$ & $P(C_{j-1} > \tfrac{k+j}{2}) + \tfrac{1}{2} P(C_{j-1} = \tfrac{k+j}{2})$
      & $P(C_{j-1} < \tfrac{k+j}{2}) + \tfrac{1}{2} P(C_{j-1} = \tfrac{k+j}{2})$  \\[3pt]
 \hline
\end{tabular}
\end{center}
\end{table}

\noindent Agent $i$ can compute $p(V,C_j \mid A_1, \cdots\!, A_j)$ as follows:
\begin{equation}
        \begin{split}
       P(V=v,\,& C_{j}=c \mid A_1, \cdots\!, A_{j}) ={}\\ &P(V=v, C_{j-1}=c \mid A_1, \cdots\!, A_{j-1})\, P(X_{j}= L \mid A_1, \cdots\!, A_{j}) + {}\\ &P(V= v,C_{j-1}=c-1 \mid A_1,\cdots\!, A_{j-1})\, P(X_{j}= H \mid A_1, \cdots\!, A_j)
        \end{split}
\end{equation}     

    When it is agent $i$'s turn to act, it will know $P(V=v, C_{i-1}=c \mid A_1, \cdots\!, A_{i-1})$ and will have its own signal $X_i$ to calculate $ P(V=v,C_i \mid C_{i-1},X_i)$. Since $V$ and $C_i$ are conditionally independent given $C_{i-1}$ and $X_i$, we have:
    \begin{equation}
    \begin{split}
     P(V=v,C_i \mid C_{i-1},X_i) = P(V=v \mid C_{i-1} ,X_i) P(C_i \mid C_{i-1} ,X_i)
    \end{split}
    \end{equation} 
    where 
    \begin{equation}
    \begin{split}
        P(V=v \mid C_{i-1}, X_i) \propto P(V=v \mid C_{i-1})P(X_i \mid V = v,C_{i-1}) \label{eq:pV_given_Ci_Xi}
    \end{split}
    \end{equation} 
    
\begin{equation}
  P(X_i\!=\!x \mid V\!=\!v, C_{i-1}) = P(X_i\!=\!x \mid V\!=\!v) =
  \begin{cases}
       p & \text{if } (v,x) \in \{(1,H), (0, L)\} \\
   1\!-\!p & \text{otherwise}
\end{cases}
\end{equation}
\noindent and 
\begin{equation}
P(C_i=c \mid C_{i-1}, X_i=x) = 
 \begin{cases}
    P(C_i = c\!-\!1) & \text{if } x = H \\
    P(C_i = c) & \text{if } x = L
\end{cases}
\end{equation}

Finally, as agent $i$ now has a probability distribution over $C_i$, it can choose between the options $A$ and $R$ by comparing $p_1 = P(C_i > \frac{k+i}{2} \mid \cdots)$, $p_2 = P(C_i < \frac{k+i}{2} \mid \cdots)$ and $p_3 = P(C_i = \frac{k+i}{2} \mid \cdots)$. For deterministic choice, if $p_1 > p_2$ it chooses $A$, and if $p_2 > p_1$ it chooses $R$. Otherwise it tosses an evenly weighted coin to choose between the two options. For non-deterministic choice, a random weighted choice is made with weights $p_1 + \frac{p_3}{2}$ for $A$ and $p_2 + \frac{p_3}{2}$ for $R$.

  \section{Experiment and Results}

\subsection{Deterministic Model}

\begin{figure} [!ht]
     \centering
     \begin{subfigure}[b]{0.3\textwidth}
         \centering
         \includegraphics[width=\textwidth]{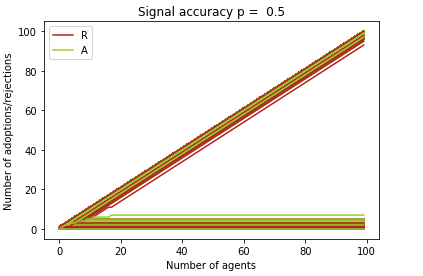}
         \caption{}
         \label{fig:model9_a}
     \end{subfigure}
     \hfill
     \begin{subfigure}[b]{0.31\textwidth}
         \centering
         \includegraphics[width=\textwidth]{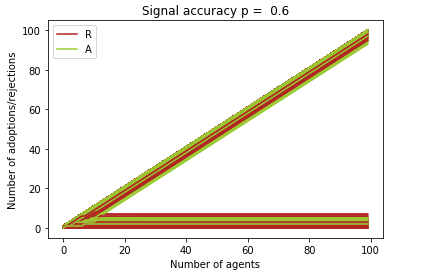}
         \caption{}
         \label{fig:model9_b}
     \end{subfigure}
     \hfill
     \begin{subfigure}[b]{0.3\textwidth}
         \centering
         \includegraphics[width=\textwidth]{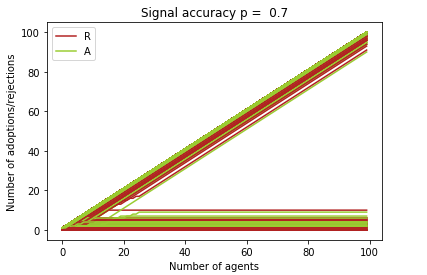}
         \caption{}
         \label{fig:model9_c}
     \end{subfigure}
     \hfill
     \begin{subfigure}[b]{0.31\textwidth}
         \centering
         \includegraphics[width=\textwidth]{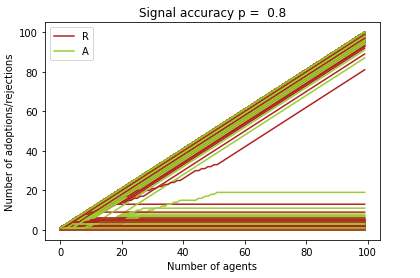}
         \caption{}
         \label{fig:model9_d}
     \end{subfigure}
     \medskip
     \begin{subfigure}[b]{0.3\textwidth}
         \centering
         \includegraphics[width=\textwidth]{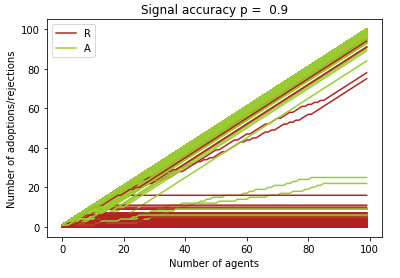}
         \caption{}
         \label{fig:model9_e}
     \end{subfigure}
        \caption{Cumulative frequencies of adopts and rejects for 100 agents over 1000 runs with $V = 1$ for each $p$ with 1 prior agent  for deterministic choice (best viewed in colour)}
\label{fig:model9}
\end{figure}

\begin{figure} [!ht]
     \centering
     \begin{subfigure}[b]{0.3\textwidth}
         \centering
         \includegraphics[width=\textwidth]{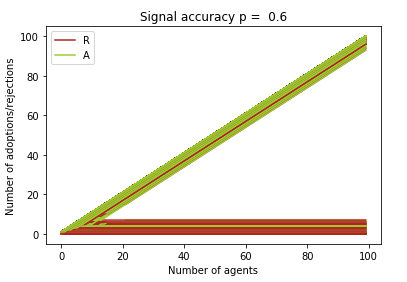}
         \caption{}
         \label{fig:model10_a}
     \end{subfigure}
     \hfill
     \begin{subfigure}[b]{0.3\textwidth}
         \centering
         \includegraphics[width=\textwidth]{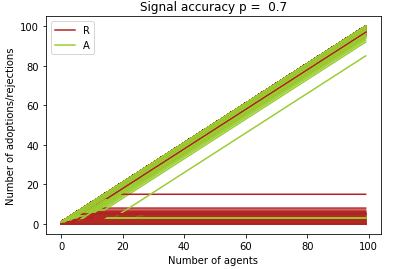}
         \caption{}
         \label{fig:model10_b}
     \end{subfigure}
     \hfill
     \begin{subfigure}[b]{0.3\textwidth}
         \centering
         \includegraphics[width=\textwidth]{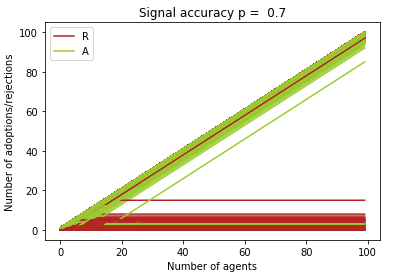}
         \caption{}
         \label{fig:model10_c}
     \end{subfigure}
     \hfill
     \begin{subfigure}[b]{0.3\textwidth}
         \centering
         \includegraphics[width=\textwidth]{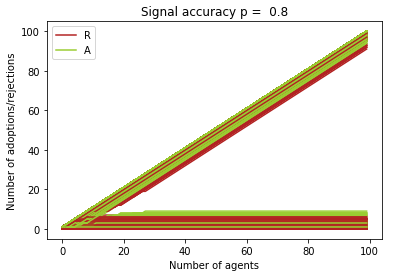}
         \caption{}
         \label{fig:model10_d}
     \end{subfigure}
     \medskip
     \begin{subfigure}[b]{0.3\textwidth}
         \centering
         \includegraphics[width=\textwidth]{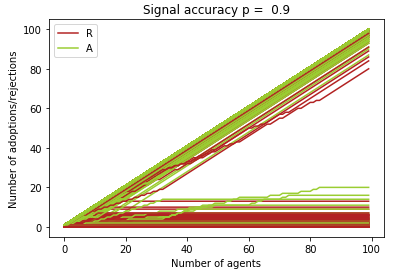}
         \caption{}
         \label{fig:model10_e}
     \end{subfigure}
        \caption{Cumulative frequencies of adopts and rejects for 100 agents over 1000 runs with $V = 1$ for each $p$ with 20 prior agents for deterministic choice (best viewed in colour)}
\label{fig:model10}
\end{figure}

\begin{figure} [!ht]
     \centering
     \begin{subfigure}[b]{0.3\textwidth}
         \centering
         \includegraphics[width=\textwidth]{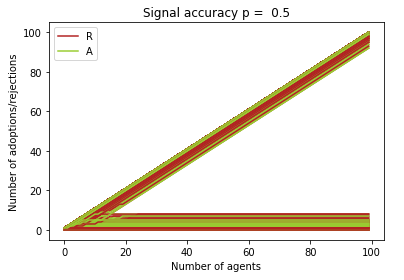}
         \caption{}
         \label{fig:model11_a}
     \end{subfigure}
     \hfill
     \begin{subfigure}[b]{0.31\textwidth}
         \centering
         \includegraphics[width=\textwidth]{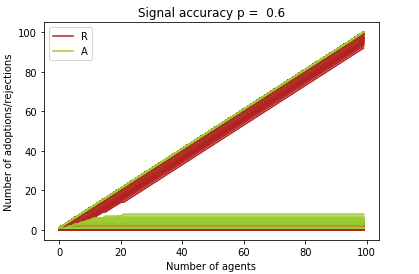}
         \caption{}
         \label{fig:model11_b}
     \end{subfigure}
     \hfill
     \begin{subfigure}[b]{0.3\textwidth}
         \centering
         \includegraphics[width=\textwidth]{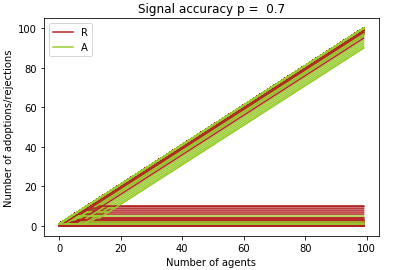}
         \caption{}
         \label{fig:model11_c}
     \end{subfigure}
     \hfill
     \begin{subfigure}[b]{0.3\textwidth}
         \centering
         \includegraphics[width=\textwidth]{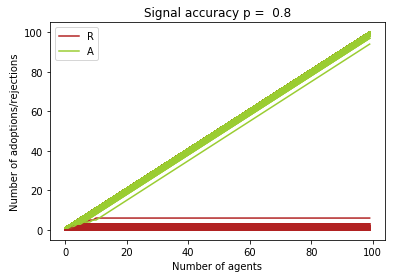}
         \caption{}
         \label{fig:model11_d}
     \end{subfigure}
     \medskip
     \begin{subfigure}[b]{0.31\textwidth}
         \centering
         \includegraphics[width=\textwidth]{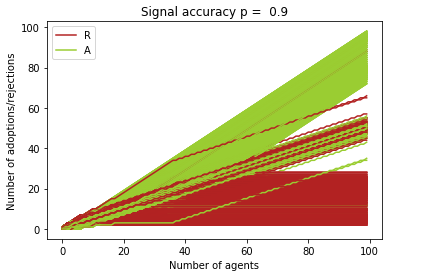}
         \caption{}
         \label{fig:model11_e}
     \end{subfigure}
        \caption{Cumulative frequencies of adopts and rejects for 100 agents over 1000 runs with $V = 1$ for each $p$ with 40 prior agents  for deterministic choice (best viewed in colour)}
\label{fig:model11}
\end{figure}

As a first step we implemented\footnote{The implementation of our model in Python can be found at https://github.com/ashalya86/Information-cascade-models.} a deterministic choice model \cite{bikhchandani_theory_1992} using our Bayesian approach. Our model uses the probability distribution $P( C_i \mid C_{i-1}, X_i)$ to choose actions.

We ran the simulation 1000 times for 100 agents for different values of $v \in \{0,1\}$, $p \in \{0.5, 0.6, 0.7, 0.8, 0.9\}$ and $k \in  \{1,20,40\}$. In a single graph we plot two lines, red and green, for each run. The green line indicates the cumulative frequency of acceptances ($A$) and the red line indicates the same for rejections ($R$). There are separate graphs for each value of $p \in \{0.5, 0.6, 0.7, 0.8, 0.9\}$. Figures \ref{fig:model9}, \ref{fig:model10} and \ref{fig:model11} show the outcome of cascades which start with 1, 20 and 40 prior agents, respectively, for $V = 1$. The results for $V = 0$ are similar to those for $V = 1$, except that the colours are swapped as $R$ is dominant when $V = 0$. It is evident from the graphs that there is always a high chance of cascades where everyone adopts when $V = 1$ and everyone rejects when $V = 0$. These graphs are similar to the results obtained from Bikhchandani's original model~\cite{bikhchandani_theory_1992}.

\subsection{Non-deterministic Model}
We wish to investigate how the deterministic model of choice in the model of Bikhchandani et al.\ impacts their results, given that non-deterministic choice has been described as more psychologically plausible~\cite{pleskac_decision_2015}. We modified our implementation to pick an action using random weighted choice. As a result, even if the most likely optimal choice is to adopt, the agent could still choose the less likely one and reject.

\begin{figure} [!ht]
     \centering
     \begin{subfigure}[b]{0.3\textwidth}
         \centering
         \includegraphics[width=\textwidth]{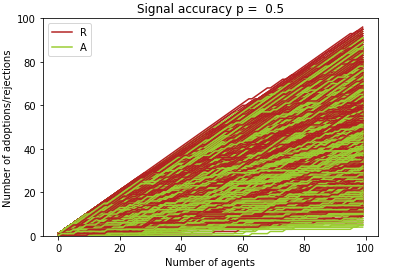}
         \caption{}
         \label{fig:model6_a}
     \end{subfigure}
     \hfill
     \begin{subfigure}[b]{0.3\textwidth}
         \centering
         \includegraphics[width=\textwidth]{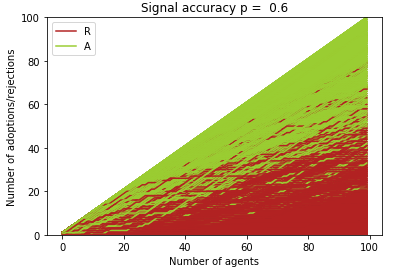}
         \caption{}
         \label{fig:model6_b}
     \end{subfigure}
     \hfill
     \begin{subfigure}[b]{0.3\textwidth}
         \centering
         \includegraphics[width=\textwidth]{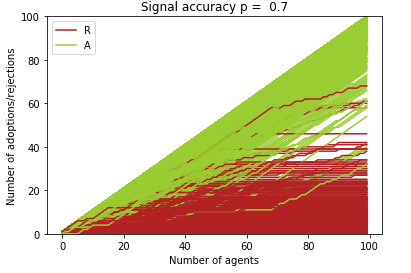}
         \caption{}
         \label{fig:model6_c}
     \end{subfigure}
     \hfill
     
     \begin{subfigure}[b]{0.315\textwidth}
         \centering
         \includegraphics[width=\textwidth]{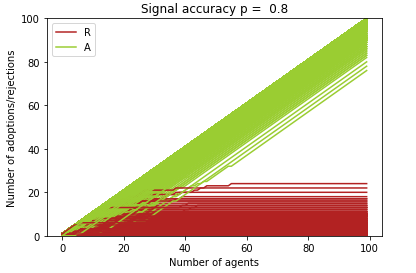}
         \caption{}
         \label{fig:model6_d}
     \end{subfigure}
     \medskip
     \begin{subfigure}[b]{0.31\textwidth}
         \centering
         \includegraphics[width=\textwidth]{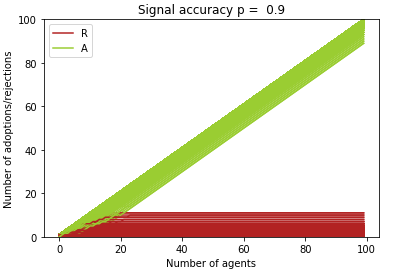}
         \caption{}
         \label{fig:model6_e}
     \end{subfigure}
        \caption{Cumulative frequencies of adopts and rejects for 100 agents over 1000 runs with $V = 1$ for each $p$ with 1 prior agent for non deterministic choice (best viewed in colour)}
\label{fig:model6}
\end{figure}

 \begin{figure} [!ht]
     \centering
     \begin{subfigure}[b]{0.3\textwidth}
         \centering
         \includegraphics[width=\textwidth]{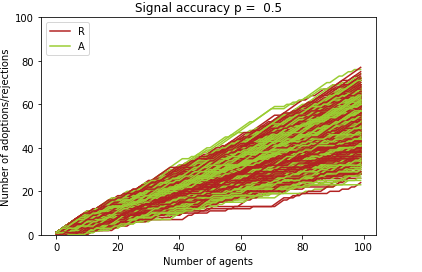}
         \caption{}
         \label{fig:model7_a}
     \end{subfigure}
     \hfill
     \begin{subfigure}[b]{0.3\textwidth}
         \centering
         \includegraphics[width=\textwidth]{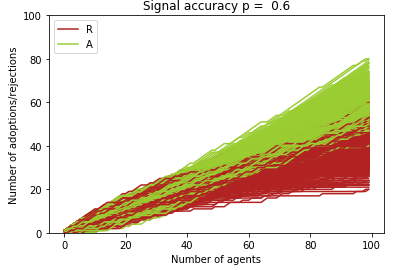}
         \caption{}
         \label{fig:model7_b}
     \end{subfigure}
     \hfill
     \begin{subfigure}[b]{0.3\textwidth}
         \centering
         \includegraphics[width=\textwidth]{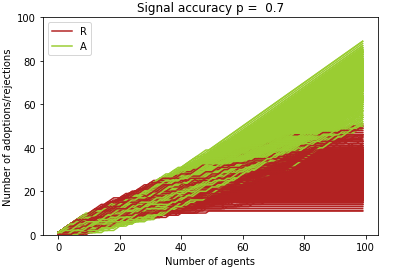}
         \caption{}
         \label{fig:model7_c}
     \end{subfigure}
     \hfill
     
     \begin{subfigure}[b]{0.31\textwidth}
         \centering
         \includegraphics[width=\textwidth]{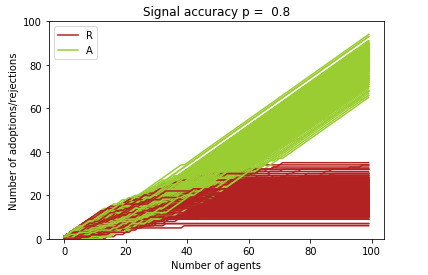}
         \caption{}
         \label{fig:model7_d}
     \end{subfigure}
     \medskip
     \begin{subfigure}[b]{0.31\textwidth}
         \centering
         \includegraphics[width=\textwidth]{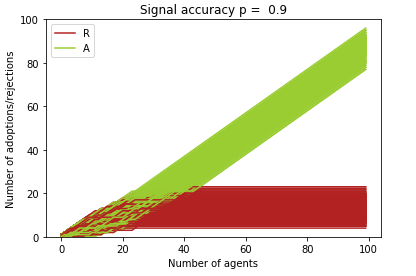}
         \caption{}
         \label{fig:model7_e}
     \end{subfigure}
        \caption{Cumulative frequencies of adopts and rejects for 100 agents over 1000 runs with $V = 1$ for each $p$ with 20 prior agents for non deterministic choice (best viewed in colour)}
\label{fig:model7}
\end{figure}

\begin{figure} [t]
     \centering
     \begin{subfigure}[b]{0.32\textwidth}
         \centering
         \includegraphics[width=\textwidth]{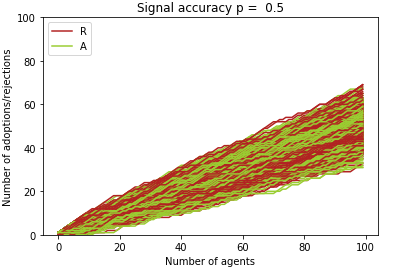}
         \caption{}
         \label{fig:model8_a}
     \end{subfigure}
     \hfill
     \begin{subfigure}[b]{0.32\textwidth}
         \centering
         \includegraphics[width=\textwidth]{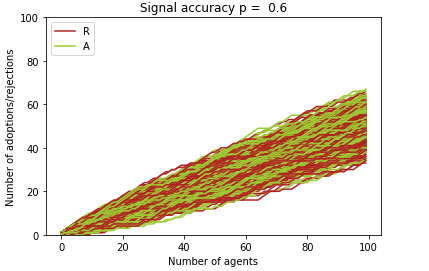}
         \caption{}
         \label{fig:model8_b}
     \end{subfigure}
     \hfill
     \begin{subfigure}[b]{0.3\textwidth}
         \centering
         \includegraphics[width=\textwidth]{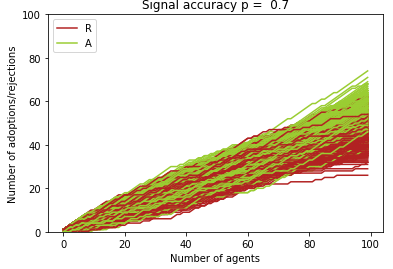}
         \caption{}
         \label{fig:model8_c}
     \end{subfigure}
     \hfill
     
     \begin{subfigure}[b]{0.31\textwidth}
         \centering
         \includegraphics[width=\textwidth]{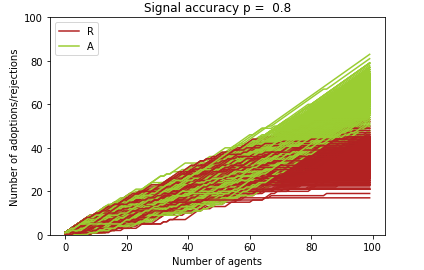}
         \caption{}
         \label{fig:model8_d}
     \end{subfigure}
     \medskip
     \begin{subfigure}[b]{0.31\textwidth}
         \centering
         \includegraphics[width=\textwidth]{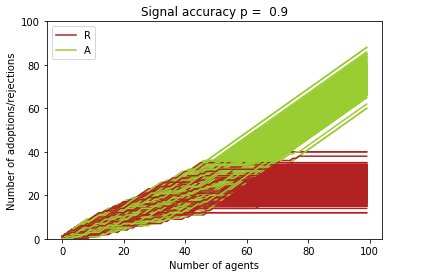}
         \caption{}
         \label{fig:model8_e}
     \end{subfigure}
        \caption{Cumulative frequencies of adopts and rejects for 100 agents over 1000 runs with $V = 1$ for each $p$ with 40 prior agents for non deterministic choice (best viewed in colour)}
\label{fig:model8}
\end{figure}

As for the deterministic choice model, in a single graph we plot cumulative frequency of acceptances ($A$) and rejections ($R$) for 1000 runs of 100 agents for each $p$. The plots of cascades which  begin with 1, 20 and 40 prior agents for $V = 1$, are seen in Figures \ref{fig:model6},  \ref{fig:model7}, \ref{fig:model8}. The results for $V = 0$ are similar to $V = 1$, except that the colours are swapped as R is dominant when $V = 0$.

Simulation results suggest that for high accuracy perception of the true value of a choice ($p \in \{0.8,0.9$\}) (Figures \ref{fig:model6_d}, \ref{fig:model6_e}, \ref{fig:model7_d} and \ref{fig:model7_e}), cascades still occur, but for lower accuracy perception, cascades are replaced by a bias towards one choice that is greater than
would be expected if only the odds of correct vs. incorrect perception were considered (Figures \ref{fig:model6_a}, \ref{fig:model6_b}, \ref{fig:model7_a} and \ref{fig:model7_b}). Although the ratio of choosing $A$s over $R$s starts with 0.6 and 0.4 for $p = 0.6$, one choice becomes increasingly dominant and it still shows the possibility of no cascade occurring. (Figures \ref{fig:model6_b} and \ref{fig:model7_b}).

According to Bikhchandani's model, the first agent's private information can be uniquely determined from its action, which makes its choice highly influential. Therefore, our model  starts by assuming there are some unobserved agents present and uses a prior distribution over $C$ and $V$.

For instance, suppose there are assumed to be 20 unobserved prior agents. Then the model creates a uniform prior for $V$ and a binomial distribution over the count of prior $H$ signals given $V$, in terms of $p$, for 20 prior agents. We then do the Bayesian inference. We obtained a significant change in cascades while plotting the cascade with different numbers of prior agents. We notice that the prior agents delay the occurrence of cascades. While the proportions of two choices have wider deviation for 1 prior agent (Figures \ref{fig:model6_b}, \ref{fig:model6_c}, \ref{fig:model6_d} and \ref{fig:model6_e}), it gradually reduces for 20 prior agents (Figures \ref{fig:model7_b}, \ref{fig:model7_c}, \ref{fig:model7_d} and \ref{fig:model7_e}) and 30 (Figures \ref{fig:model8_b}, \ref{fig:model8_c}, \ref{fig:model8_d} and \ref{fig:model8_e})

\section{Conclusion}
An information cascade happens when people observe the actions of their predecessors and try to follow these observations regardless of their own private information. We presented a full Bayesian account for the model of Bikhchandani et al.~\cite{bikhchandani_theory_1992}. We maintain a probability distribution over $V$ and the (unknown) counts of High signals received by the agents. Rather than always choosing the most likely option, agents make a weighted choice between Adopt and Reject. We do not assume that the first agent that was observed was the first to consider the fad. Instead, we incorporate prior knowledge of unobserved agents. This means that first (observed) agent’s choice is less dominant than in the earlier model. Our findings show that prior agents delay the occurrence of cascades. Furthermore, in contrast to the predictions of Bikhchandani et al., our results show that cascades will not necessarily occur. The graphs obtained show that, for lower accuracy perception, cascades occur with much less probability. However, as $p$ gets high, there is a high chance of cascades.

While people may not be good at Bayesian reasoning, as used in our model, we believe that appropriate software using our model could support users to assess the evidence from others’ choices, including the possible presence of unobserved agents. This could help to reduce the likelihood of cascades.


%
%
%
 \bibliographystyle{splncs04}
 \bibliography{coine2021}

\end{document}